\documentclass[%
 reprint,
superscriptaddress,
 amsmath,amssymb,
 aps,
]{revtex4-2}

\usepackage{graphicx}
\usepackage{dcolumn}
\usepackage{color}
\usepackage{bm}

\newcommand{\SLAC}{SLAC National Accelerator Laboratory, Menlo Park, CA 94025}
\newcommand{\AP}{Department of Applied Physics, Stanford University, Stanford, CA 94305}
\newcommand{\PULSE}{Stanford PULSE Institute, SLAC National Accelerator Laboratory, Menlo Park, CA 94025, USA}

\begin{document}

\preprint{APS/123-QED}

\title{Spectrotemporal shaping of attosecond x-ray pulses with a fresh-slice free-electron laser}

\author{River~R.~Robles}
\email{riverr@stanford.edu}
\affiliation{\SLAC}
\affiliation{\AP}
\affiliation{\PULSE}

\author{Kirk~A.~Larsen}
\email{larsenk@stanford.edu}
\affiliation{\SLAC}
\affiliation{\PULSE}

\author{David~Cesar}
\affiliation{\SLAC}

\author{Taran~Driver}
\affiliation{\SLAC}
\affiliation{\PULSE}

\author{Joseph~Duris}
\affiliation{\SLAC}

\author{Paris~Franz}
\affiliation{\SLAC}
\affiliation{\AP}
\affiliation{\PULSE}

\author{Douglas~Garratt}
\affiliation{\SLAC}
\affiliation{\PULSE}

\author{Veronica~Guo}
\affiliation{\SLAC}
\affiliation{\AP}
\affiliation{\PULSE}

\author{Gabriel~Just}
\affiliation{\SLAC}

\author{Randy~Lemons}
\affiliation{\SLAC}

\author{Ming-Fu~Lin}
\affiliation{\SLAC}
\affiliation{\PULSE}

\author{Razib Obaid}
\affiliation{\SLAC}
\affiliation{\PULSE}

\author{Nicholas~Sudar}
\affiliation{\SLAC}

\author{Jun~Wang}
\affiliation{\SLAC}
\affiliation{\AP}
\affiliation{\PULSE}

\author{Zhen~Zhang}
\affiliation{\SLAC}

\author{James~Cryan}
\email{jcryan@slac.stanford.edu}
\affiliation{\SLAC}
\affiliation{\PULSE}

\author{Agostino~Marinelli}
\email{marinelli@slac.stanford.edu}
\affiliation{\SLAC}
\affiliation{\PULSE}

\date{\today}

\begin{abstract}
We propose a scheme allowing coherent shaping, i.e., controlling both the amplitude and phase, of attosecond x-ray pulses at free-electron lasers. 
We show that by seeding an FEL with a short coherent seed that overfills the amplification bandwidth, one can shape the Wigner function of the pulse by controlling the undulator taper profile. 
The examples of controllable pulse pairs and trains, as well as isolated spectrotemporally shaped pulses with very broad bandwidths are examined in detail. Existing attosecond XFELs can achieve these experimental conditions in a two-stage cascade, in which the seed is generated by a short current spike in an electron bunch and shaped in an unspoiled region within the same bunch. 
We experimentally demonstrate the production and control of phase-stable pulse trains using this method at the Linac Coherent Light Source II.

\end{abstract}

\maketitle

Optical lasers have had a far reaching impact on science and society. Key to their broad success has been the development of shaping methods allowing the production of waveforms with programmable amplitude, phase, frequency, and pulse format \cite{weiner1988high,hillegas1994femtosecond,weiner2000femtosecond,shim2006femtosecond,weiner2011ultrafast}. Such control has many different forms and applications. Control over the time-dependent phase of an optical pulse enables, for example, chirped pulse amplification, whose scientific impact is so significant that it was recently recognized with a Nobel prize \cite{maine1988generation}. Shaping of laser pulses has fundamental scientific applications in coherent control, in which engineering the precise amplitude and phase of an optical pulse is used to exert control over the dynamics of a quantum system \cite{brumer1986control,shi1988optimal,kosloff1989wavepacket,goswami2003optical,ohmori2009wave}. 

Free-electron lasers (FELs) have extended our access to high intensity coherent radiation sources into the extreme ultraviolet (EUV) and x-ray regions \cite{bonifacio1984collective,emma2010first,allaria2012highly}. EUV FELs benefit from conventional laser technology since they can be seeded by external lasers in harmonic generating schemes \cite{yu2000high,yu2003first,xiang2009echo,xiang2010demonstration,zhao2012first}. Externally seeded FELs transfer the coherence in the seed laser to the EUV pulse. Combined with the inherent flexibility of FEL systems, seeding enables a variety of shaping methods \cite{gauthier2015spectrotemporal,maroju2020attosecond}. To date, several such possibilities have been demonstrated: the production of phase correlated two-color pulses \cite{prince2016coherent}, stable attosecond pulse trains \cite{maroju2023attosecond}, and chirped pulse amplification \cite{gauthier2016chirped}. 

X-ray FELs typically operate in the self amplified spontaneous emission (SASE) mode, in which the FEL interaction is seeded by shot noise in the electron beam. As such, SASE light is typically temporally incoherent, composed of a number of uncorrelated temporal and spectral spikes. Shaping of SASE x-ray pulses has been achieved by shaping the phase-space of the electron bunch, resulting in multicolor FELs \cite{marinelli2015high, lutman2016fresh}, and pulses of variable duration and structure
\cite{ding2015generating,marinelli2016optical, duris2021controllable}, down to the attosecond regime \cite{duris2020tunable, prat2023x, huang2017generating}, and by control over the undulator line \cite{prat2024experimental}.

 
In this letter, we propose a coherent spectrotemporal shaping method for attosecond XFEL pulses. By coherent shaping, we mean exerting control over both the amplitude and phase of the x-ray pulse.
Leveraging the interplay between slippage and the finite FEL gain bandwidth, we show that the bandwidth of an isolated attosecond pulse can be manipulated to produce a wide variety of output pulses with different spectrotemporal characteristics.
Unlike other proposed methods at x-ray energies, our approach enables both amplitude and phase shaping.
Since our method utilizes a spectrally coherent seed -- an initial isolated attosecond pulse -- the resulting shaped pulses intrinsically inherit that coherence. This method enables a variety of new experimental techniques, ranging from bandwidth broadening for impulsive stimulated X-ray Raman scattering \cite{o2020electronic}, to the realization of advanced beam microbunching schemes for the production of shorter attosecond pulses \cite{tanaka2015proposal}. Additionally, the proposed method enables the  generation and  shaping of electronic wavepackets beyond the impulsive limit \cite{li2022attosecond}, providing a path to understanding the coupling between electronic and nuclear dynamics in complex molecules. 

\begin{figure}[h!]
    \centering
    \includegraphics[width=0.85\columnwidth]{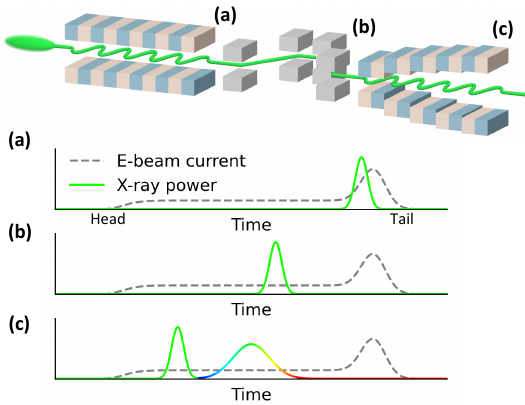}
    \caption{An illustration of the pulse shaping scheme. An isolated attosecond pulse produced in one undulator stage (a) is slipped onto a fresh part of the electron beam in a magnetic chicane (b). Varying the taper of the undulator in this second stage enables time-dependent control of the e-beam microbunching frequency and subsequent x-ray emission (c). In (a)-(c), the head of the beam is at the left.}
    \label{fig:cartoon}
\end{figure}

Figure~\ref{fig:cartoon} illustrates the proposed method. First, a current spike in the electron beam produces an isolated attosecond pulse in an initial undulator stage (a). We will refer to this first pulse as the ``seed pulse" throughout the paper. This is the standard method for producing attosecond pulses at XFELs and can be achieved with a variety of beam manipulation techniques \cite{duris2020tunable,zhang2020experimental,cesar2021electron,duris2020tunable,macarthur2019phase}.
The beam is then delayed with respect to the seed in a magnetic chicane (b). The seed pulse is now overlapped with a lower current ``fresh" slice, i.e. a slice that was not spoiled by the FEL process in the first stage. The attosecond pulse acts as a coherent seed for lasing in a second stage, in which by varying the taper of the undulator we can produce a second pulse with a flexible spectrotemporal structure (c). The physical mechanism behind this spectrotemporal shaping is an interplay between slippage and the finite bandwidth of the FEL interaction. At any moment, the seed imposes an energy modulation on the narrow temporal slice of the beam it overlaps with at the local undulator resonant frequency. Changing that frequency through the taper as the seed slips over different regions of the beam enables the imprinting of spectrotemporal structures in the bunch which are later inherited by the x-rays the bunch emits.

We can understand the basic relationship between the induced microbunching and the undulator taper using the one-dimensional collective variables model of the FEL. The collective variable equations can be written 
\begin{eqnarray}
    \frac{db}{d\hat{z}}&=&ip-i\kappa(\hat{z})b\label{eqn:collectivevariables_b}\\
    \frac{dp}{d\hat{z}} &=& -a-i\kappa(\hat{z})p\\
    \left(\frac{\partial}{\partial \hat{z}}+\frac{1}{2\rho}\frac{\partial}{\partial\theta}\right)a&=&b\label{eqn:collectivevariables_a}
\end{eqnarray}
where $b$ is the bunching factor, and $p$ and $a$ are the energy modulation and radiation field envelope normalized to their saturation values. Furthermore, $\hat{z}=2\rho k_uz$, where $z$ is distance along the undulator, $k_u=2\pi/\lambda_u$ the undulator wavenumber, $\lambda_u$ the undulator period, and  $\rho$ the FEL Pierce parameter. The ponderomotive phase $\theta=(k_r+k_u)z-\omega_rt$ can be interpreted as the position along the electron bunch multiplied by the bunching wavenumber $k_r+k_u\simeq k_r$. Finally, $\kappa(\hat{z})=\frac{1}{\rho}\frac{K_0^2}{2+K_0^2}\Delta(\hat{z})$ is the normalized undulator detuning, where the peak undulator strength parameter $K$ has been written $K(\hat{z})=K_0(1+\Delta(\hat{z}))$. $K_0$ is the reference strength used in the calculation of the reference resonant wavelength, and $\Delta$ is the relative change in $K$ from that reference $K_0$.

Equations~\eqref{eqn:collectivevariables_b}-\eqref{eqn:collectivevariables_a} can be solved analytically for linear tapers $\kappa(\hat{z})=\kappa_0+\kappa_1\hat{z}$ \cite{baxevanis2018time}. We consider a general taper $\kappa(\hat{z})$, but suppose that the field interacts weakly with the beam (either due to a significant taper, or a very broad bandwidth seed whose spectral content is mostly non-resonant). In that case, the seed effectively slips over the beam without perturbation from the interaction: $a(\theta,\hat{z})=a(\theta-\frac{\hat{z}}{2\rho},0)\equiv a_0(\theta-\frac{\hat{z}}{2\rho})$ where $a_0(\theta)$ is the initial field. For a short seed field we can approximate $a_0(\theta)\simeq A_0\delta(\theta)$. The resulting bunching is approximately
\begin{eqnarray}
    b(\theta,\hat{z}) = -2\rho iA_0(\hat{z}-2\rho\theta)H(\hat{z}-2\rho\theta)e^{i\int_{\hat{z}}^{2\rho\theta}\kappa(\hat{z}'')d\hat{z}''}.
\end{eqnarray}
where $H(x)$ is the Heaviside function.
The $\theta$-dependent bunching phase, $\phi=\int_{\hat{z}}^{2\rho\theta}\kappa(x)dx$, encodes the time-frequency chirp induced by the undulator taper. 
This equation tells us that a given taper profile induces a microbunching spectrotemporal structure of the same qualitative shape, translating the local resonance of the undulators into a local modulation frequency mediated by slippage.
For example, a linear taper $\kappa(\hat{z})=\kappa_1\hat{z}$ leads to a quadratic phase shift $\phi=2\rho^2\kappa_1\left[(k_rz-\omega_rt)^2+2k_uz(k_rz-\omega_rt)\right]$, implying a linear chirp on the bunching. We have assumed a $\delta$-function seed to illustrate the basic physics, more generally these effects are observed when the seed pulse is shorter than the FEL cooperation length in the second stage, or equivalently the seed bandwidth is broader than the FEL amplification bandwidth.

This analysis illustrates the basic shaping concept, but is oversimplified in several ways. For one, the FEL amplification bandwidth shrinks as the beam propagates. In particular, it diverges at $z=0$, implying that no finite bandwidth seed pulse can be broader than the FEL bandwidth for all $z$. Furthermore, we have ignored diffraction, which affects the field strength observed by the beam at different points in the taper. We have also ignored the effect of the FEL interaction on the seed, which can introduce changes to the seed field and the beam itself. We will study these nuances numerically.

We now explore more realistic scenarios using the GENESIS 1.3 v4 code \cite{reiche1999genesis,genesis4github}, a common tool for performing 3D FEL simulations. In the simulations, an electron bunch with constant current is seeded by a 560 eV pulse with 500 attosecond full-width at half-maximum duration (see supplementary information for more simulation details). The undulator lattice is the LCLS-II soft x-ray line, with 87-period segments with a 3.9 cm period. We focus on the seeded physics of the second undulator stage, since the production of isolated attosecond pulses (our seed pulse) has been studied extensively in the past. 

\begin{figure}[h!]
    \centering
    \includegraphics[width=0.85\columnwidth]{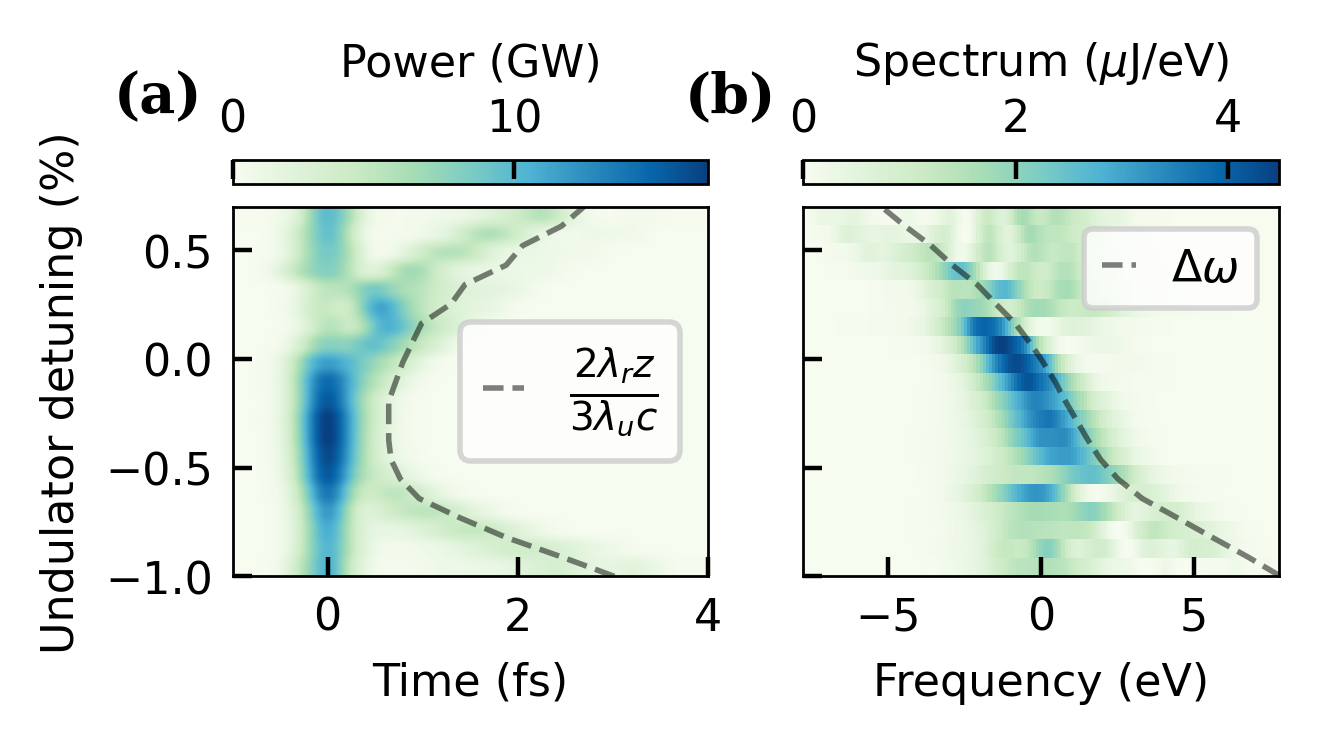}
    \caption{Simulations of a short seed pulse in a detuned undulator. (a) The power profile after energy doubling. (b) The spectrum after energy doubling. $\Delta\omega$ is defined in the text. The frequency axis is measured relative to 560 eV.}
    \label{fig:pure_frequency_pulling}
\end{figure}

The simplest taper is a constant undulator strength which may not be resonant with the central frequency of the seed, $K=K_0(1+\Delta)$. Figure~\ref{fig:pure_frequency_pulling} shows (a) the power profile and (b) the output spectrum as a function of the relative detuning of the undulators. For each detuning we show the power and spectrum at the point at which the pulse energy has doubled -- this choice is made to highlight the characteristics of experimentally interesting working points. Generally, a second pulse is emitted shortly after the seed with a shifted frequency. This basic process is called ``frequency pulling", which has been studied previously in the asymptotic, long undulator limit \cite{allaria2010tunability,allaria2011experimental,mirian2021generation}. Here we are interested in a transient regime where the two pulses coexist \cite{robles:ipac23-tupl094}. The shift of the central frequency of the second pulse from the seed is expected to be $\Delta\omega=\frac{\sigma_{\omega,s}^2}{\sigma_{\omega,s}^2+\sigma_{\omega,\text{FEL}}^2}(\omega_r-\omega_s)$ \cite{allaria2010tunability}, where $\sigma_{\omega,s}$ and $\sigma_{\omega,\text{FEL}}$ are the seed and the FEL gain bandwidths, respectively, and $\omega_r$ and $\omega_s$ are the undulator resonant frequency and the central frequency of the seed, respectively. The FEL bandwidth can be written as $\sigma_{\omega,\text{FEL}}=\omega_r\sqrt{\frac{3\sqrt{3}\rho}{k_uz}}$ (see, for example, \cite{huang2007review,pellegrini2016physics,kim2017synchrotron}), we note that it shrinks like $z^{-1/2}$. The undulator resonant frequency is $\omega_r=\frac{2\gamma^2k_uc}{1+\frac{K^2}{2}}$, where $\gamma$ is the beam Lorentz factor.

The time delay between the two pulses derives from their different group velocities. For large detunings, the seed is not affected by its interaction with the beam and slips at the speed of light. The second pulse, on the other hand, is produced by the exponential FEL instability, and moves at the FEL group velocity $v_g=c\left(1-\frac{2\lambda_r}{3\lambda_u}\right)$. After length $z$, a delay builds up of value $\Delta t=\frac{2\lambda_rz}{3\lambda_uc}$. This estimate is shown in panel (a). 
For sufficiently small detunings the seed pulse can give rise to superradiant behavior \cite{bonifacio1985superradiant,bonifacio1989superradiance,bonifacio1991superradiant,giannessi2005nonlinear}, a case which has been examined in detail in \cite{franz2024,robles2024three}. 

\begin{figure}[h!]
    \centering
    \includegraphics[width=0.9\columnwidth]{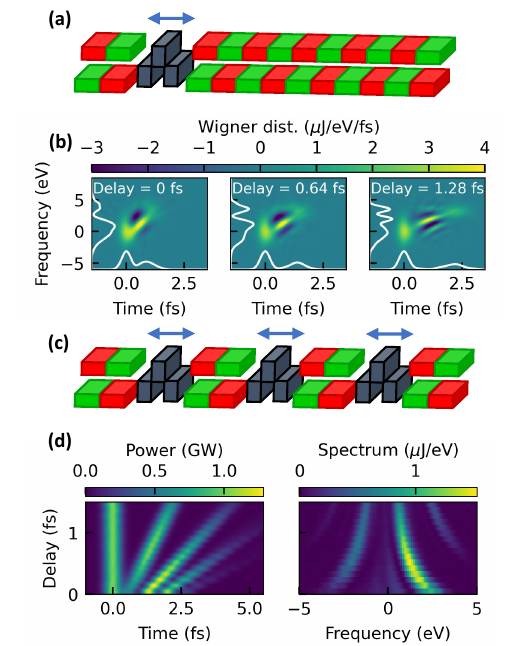}
    \caption{Simulations of a detuned undulator with intermittent delay stages. (a) shows a single delay stage after a short modulating section. (b) shows the Wigner distributions of the pulse for 0, 1, and 2 delay undulator sections. (c) shows periodic delay stages. (d) shows the power and spectrum of the pulse while scanning the periodic delay. The frequency axis is measured relative to 560 eV.}
    \label{fig:pairs_and_trains}
\end{figure}

Delay stages can enable more flexible control of the time delay. Such delay stages may be an undulator detuned far from the resonance, or a magnetic chicane between undulator segments. Figure~\ref{fig:pairs_and_trains} shows two applications of delays. In panels (a) and (b), one 87-period section is detuned $-0.7\%$ in $K$ from the seed resonance, followed by $0$, $1$, or $2$ undulators in a far detuned delay stage mode. The subsequent undulators are set back to the original detuning. Panel (b) shows the Wigner distribution of the radiation after energy doubling for the three delay options. The primary change is delaying of the second pulse by roughly $87\times \frac{\lambda_r}{c}\simeq 0.64$ fs per additional delay stage without changing its central frequency. Panels (c) and (d) extend this concept to periodic delays after every undulator section, leading to the emission of a train of phase-locked attosecond pulses. We scan the delay per stage from $0$ to $1.5$ fs. The power profile is composed of a train of attosecond pulses, and the spectra show fringes whose periodic spacing decreases as the time spacing increases, as expected for multiple pulses with mutual phase stability. The pulse train profile is made cleaner by introducing a $\pi$ phase shift between undulator segments and linearly tapering the individual segments, as discussed in more detail in Supplementary Section I. We note that using too large of a delay in these schemes can wash out the microbunching we are trying to generate. The value of that limit depends on the gain length in the modulating sections and the seed field strength.

The method described thus far generally yields a secondary shaped pulse (or train of pulses) while the seed pulse is largely unperturbed. To generate isolated shaped pulses, we can leverage harmonic upconversion of the induced microbunching, and radiate at a harmonic of the seed (the seed pulse could be removed by a frequency filter or used in a two-pulse experiment \cite{guo2024,li2024attosecond}). For the simulations below, we assume a Fourier transform limited 280 eV seed pulse with 600 attosecond FWHM duration. 

\begin{figure}[h!]
    \centering
    \includegraphics[width=0.9\columnwidth]{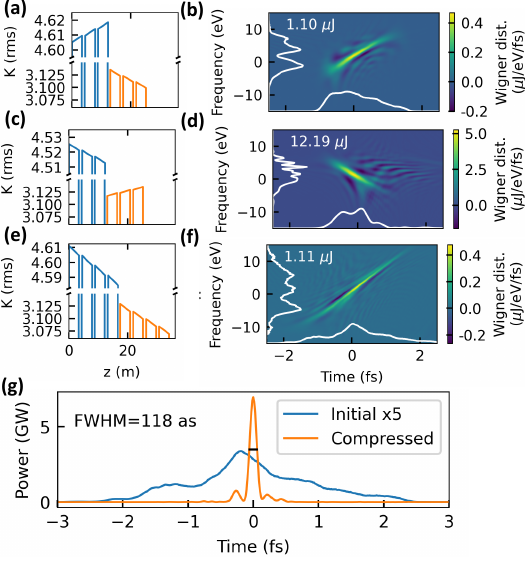}
    \caption{Simulations of linearly tapered undulators with second harmonic upconversion. (b), (d), and (f) show the Wigner distribution of the second harmonic pulse for the undulator tapers shown in (a), (c), and (e), respectively. (g) shows the effect of an ideal pulse compressor on the pulse from (e)-(f). The frequency axis is measured relative to 560 eV.}
    \label{fig:harmonic_chirps}
\end{figure}

Figure~\ref{fig:harmonic_chirps} shows three examples. In each case, a linearly tapered undulator imposes roughly linearly chirped microbunching. Following that, a second linearly tapered undulator tuned to the second harmonic of the first prompts lasing at the second harmonic, 
producing a linearly chirped pulse. Figure~\ref{fig:harmonic_chirps}(c) and (d) show a case leading to a negative linear chirp, while panels (a)-(b) and (e)-(f) show cases leading to a positive chirp. The fact that cases (a) and (e) yield the same direction of chirp despite having different taper directions derives from the fact that the FEL gain bandwidth $\sigma_{\omega,\text{FEL}}(z)$ diverges at $z=0$. That means that any finite bandwidth seed is narrower than the amplification bandwidth at the beginning of the undulator. This leads to the local pulled frequency (given before as $\Delta\omega\simeq \frac{\sigma_{\omega,s}^2}{\sigma_{\omega,s}^2+\sigma_{\omega,\text{FEL}}^2}(\omega_r-\omega_s)$) having a different shape and even a different sign slope than the taper profile. 

Panels (e) and (f) highlight the ability to generate very large spectral bandwidths with this approach. The ability to generate larger bandwidths than are inherent to the seed stems from FEL gain at the edges of the seed spectrum, similar to spectral broadening in optical lasers \cite{nisoli1996generation,nisoli1997compression,travers2019high}. In panel (g) we show the power after passage through an ideal optical compressor, which we simulate by applying a quadratic phase to the field in the frequency domain. We find that the very broad bandwidth linearly chirped pulse is compressible to $120$ as FWHM duration, a factor of four shorter than previously observed with XFELs \cite{duris2020tunable} and five times shorter than the seed. Physically, this compression could be accomplished with consecutive dispersive reflections \cite{chapman2002x,fujiki2009compression,hrdy2013possibility}. Such a scheme requires high-reflectivity, broad bandwidth dispersive soft x-ray optics which are an area of active research \cite{bajt2012high,de2015phase,guggenmos2013aperiodic,hofstetter2011attosecond,huang2017spectral}.

\begin{figure*}[htb!]
    \centering
    \includegraphics[width=0.9\linewidth]{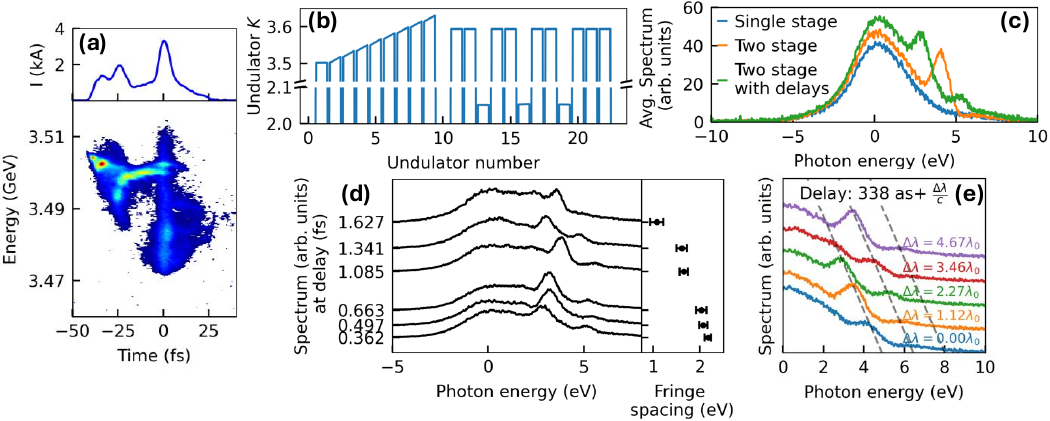}
    \caption{Experimental demonstration of phase-locked pulse train generation and control. (a) The measured beam phase space. (b) An example undulator taper profile with periodic delay stages. (c) Average spectra from the first stage, the second stage with no delays (frequency pulling), and the second stage with delays. (d) Average spectrum versus delay in the coarse scan, alongside the spectral fringe spacing. (e) Average spectrum versus delay in the fine scan, with guiding lines tracking the fringes. Photon energy axes are all relative to 400 eV.}
    \label{fig:experiment}
\end{figure*}

We conclude with an experimental demonstration of the production and control of phase-stable spectral fringes at the LCLS-II. A 3.5 GeV electron beam was shaped by cathode laser stacking, analogously to \cite{zhang2020experimental}, to generate a femtosecond-scale current spike on top of a longer, flat current profile as shown in Figure~\ref{fig:experiment}(a). We note that the few-fs temporal resolution of the x-band transverse deflecting cavity (XTCAV) is insufficient to properly resolve the short current spike at $0$ fs. The first 9 undulators were positively tapered (see panel (b)), allowing the chirped current spike to emit an isolated pulse with 400 eV central photon energy, $5$ $\mu$J average pulse energy, and an average spectrum shown by the blue curve in panel (c) measured using a Fresnel Zone Plate spectrometer \cite{larsen2023compact}. 
A magnetic chicane located at undulator index 10 delayed the beam 10 fs, overlapping this first pulse with fresh electrons. The remaining undulators were first tuned to emit a second pulse frequency pulled from the seed by 4 eV, as shown by the average spectrum plotted in orange in panel (c). Subsequently, we detuned every third undulator in the second stage to periodically delay the beam. Panel (b) of Figure~\ref{fig:experiment} shows an example taper profile, where every third undulator is set to $K=2.05$. The delay induced by each undulator is the slippage length, $N_u\lambda_\text{detuned}$, where $N_u=87$ is the number of periods per undulator section, and $\lambda_\text{detuned}$ is the resonant wavelength in the detuned sections. After moving to the periodic delay mode, the pulled part of the spectrum split into fringes, as in the green curve in panel (c). The fact that these fringes survive after averaging hundreds of shots is evidence of the presence of a pulse train with shot-to-shot mutual phase stability.

The pulse train can be controlled by changing the delay undulator $K$ values. First, we scanned the delay added per detuned section coarsely from 360 as to 1.6 fs, which is expected to change the temporal spacing between pulses in the train. That increase in the pulse-to-pulse spacing is reflected in a decrease in the spacing between the spectral fringes, as shown on the right side of panel (d) estimated by gaussian fitting to the spectra (see supplementary information for details). Next we set the delay to 338 as and scanned finely over just a few radiation periods $\lambda_0/c$, where $\lambda_0$ is the wavelength of the 400 eV seed pulse. Such small changes enable control over the phase shift between adjacent pulses without significantly changing the pulse-to-pulse spacing. A shift in the pulse-to-pulse phase shift in a periodic train manifests as a shift in the location of the spectral fringes, which we observe in panel (e). 

In conclusion, we have presented a versatile method for the shaping of attosecond x-ray pulses at free-electron lasers. The method leverages the interplay of slippage and the finite FEL gain bandwidth in order to impose flexible, time-dependent microbunching on the beam that can later lase producing similarly structured pulses. By combining detuned undulators with delay stages we have shown the possibility to generate phase-locked pulse pairs and pulse trains with controllable temporal spacing and color separation. By leveraging harmonic microbunching, frequency-isolated shaped pulses can be produced and used on their own or combined with the initial attosecond pulse for pump-probe experiments. The exponential seeded FEL gain process allows us to generate very large bandwidths in this way. Finally, we have demonstrated the production and control of pulse trains with pulse-to-pulse phase stability at the LCLS-II using periodic delay undulators. 

This work was supported by the U.S. Department of Energy, Office of Science, Office of Basic Energy Sciences under
Contract No. DE-AC02-76SF00515, and by the U.S. Department of Energy, Office of Science, Office of Basic Energy Sciences Accelerator and Detector Research Program. R. R. R. acknowledges the support of the Stanford Graduate Fellowship and the Robert H. Siemann Fellowship. The authors thank Zhirong Huang for helpful discussions.

\bibliography{apssamp}

\end{document}


\title[Spectrotemporal shaping of attosecond x-ray pulses with a fresh-slice free-electron laser]{Supplementary Information: Spectrotemporal shaping of attosecond x-ray pulses with a fresh-slice free-electron laser}

\author{River~R.~Robles}
\email{riverr@stanford.edu}
\affiliation{\SLAC}
\affiliation{\AP}
\affiliation{\PULSE}

\author{Kirk~A.~Larsen}
\email{larsenk@stanford.edu}
\affiliation{\SLAC}
\affiliation{\PULSE}

\author{David~Cesar}
\affiliation{\SLAC}

\author{Taran~Driver}
\affiliation{\SLAC}
\affiliation{\PULSE}

\author{Joseph~Duris}
\affiliation{\SLAC}

\author{Paris~Franz}
\affiliation{\SLAC}
\affiliation{\AP}
\affiliation{\PULSE}

\author{Douglas~Garratt}
\affiliation{\SLAC}
\affiliation{\PULSE}

\author{Veronica~Guo}
\affiliation{\SLAC}
\affiliation{\AP}
\affiliation{\PULSE}

\author{Gabriel~Just}
\affiliation{\SLAC}

\author{Randy~Lemons}
\affiliation{\SLAC}

\author{Ming-Fu~Lin}
\affiliation{\SLAC}
\affiliation{\PULSE}

\author{Razib Obaid}
\affiliation{\SLAC}
\affiliation{\PULSE}

\author{Nicholas~Sudar}
\affiliation{\SLAC}

\author{Jun~Wang}
\affiliation{\SLAC}
\affiliation{\AP}
\affiliation{\PULSE}

\author{Zhen~Zhang}
\affiliation{\SLAC}

\author{James~Cryan}
\email{jcryan@slac.stanford.edu}
\affiliation{\SLAC}
\affiliation{\PULSE}

\author{Agostino~Marinelli}
\email{marinelli@slac.stanford.edu}
\affiliation{\SLAC}
\affiliation{\PULSE}

\date{\today}

\maketitle

\section{Simulation details}

The relevant parameters for all simulations are shown in Table~\ref{tab:sim_parameters}. The values without parentheses are relevant for the multi-pulse simulations (Figures 2 and 3 in the main text) except for the peak power for the pulse trains portion of Figure 3, for which the power is given in brackets in the table. The values in parentheses are relevant for Figure 4.

\begin{table}[h!]
    \centering
    \begin{tabular}{c|c}
        \hline
        Beam parameter & Value  \\
        \hline
        Current & 2 kA\\
        Average beam size & 20 $\mu$m\\
        Normalized emittance & 0.4 $\mu$m\\
        Energy & 5 GeV \\
        Energy spread & 2 MeV \\
        \hline
        Undulator parameter & Value\\
        \hline
        Period & 3.9 cm \\
        Periods per segment & 87 \\
        \hline
        Seed field parameter & Value \\
        \hline 
        Central frequency & 560 (280) eV\\
        FWHM power duration & 0.5 (0.6) fs\\
        Peak power & 10 [1] GW\\
        Spot size $w_0$ \cite{siegman1986lasers} & 70.7 (100) $\mu$m
    \end{tabular}
    \caption{Parameters for simulations. Meanings of parentheses and brackets are explained in the text.}
    \label{tab:sim_parameters}
\end{table}

In the pulse pair simulations with delays (Figure 3), the delay stages are undulator segments detuned 4\% in $K$. In the pulse train simulations, the delay stages are represented by phase shifters between segments. To clean up the temporal profile of the train we applied, in addition to the macroscopic delay, a $\pi$ phase shift at each phase shifter. Furthermore, we found that the undulator segments in the LCLS-II lattice were long enough that sufficient gain occurred to distort the temporal profile. To clean up these effects, we linearly tapered each individual segment so as to reduce the coupling between the beam and the radiation, effectively shortening the segment.

\section{Experiment details and data analysis}

The experiment was performed at the LCLS-II facility. The usual $\simeq 16$ ps long, gaussian cathode laser pulse was shaped by the addition of a shorter, $3$ ps long pulse, leading to analogous seeding of the microbunching instability downstream as reported in \cite{zhang2020experimental}. The energy modulation generated in the three linac sections and two bunch compressors was finally compressed to a fs, 10 kA scale current spike by introducing an orbit bump in the final dogleg prior to the LCLS-II undulators. In the drift from the dogleg to the undulators, longitudinal space charge forces generated a 35 MeV peak to peak linear chirp within the spike. We compensated that roughly linear chirp with the positive undulator taper. 

The spectrum at the output of the second stage was measured using a Fresnel Zone Plate spectrometer \cite{larsen2023compact} in the time-resolved AMO (TMO) experimental end station. 

\subsection{Data analysis in the coarse delay scan}

\begin{figure}[h!]
    \centering
    \includegraphics[width=0.9\linewidth]{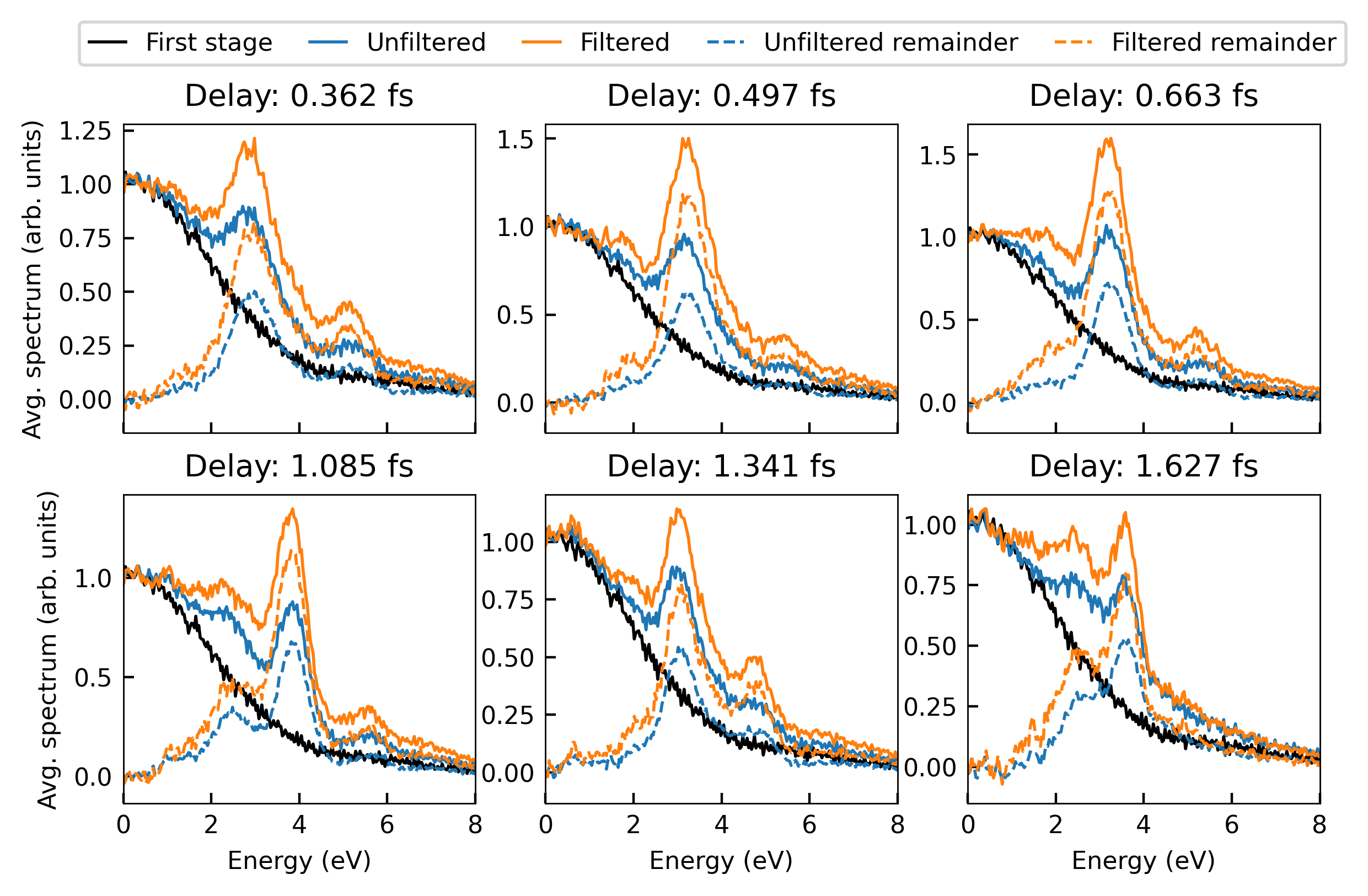}
    \caption{Average filtered and unfiltered spectra for the six delay setpoints in the coarse delay scan. The dashed lines are the background-subtracted difference spectra to be used for further analysis.}
    \label{fig:spec_bgd_removal}
\end{figure}

To estimate the fringe spacing we used averaged spectra, and used the bootstrapping method to estimate the uncertainty in the fringe spacings. We worked with two different sets of average spectra. The first was an average over all shots in each dataset, about 500 each, with no filtering. We also studied a reduced dataset consisting of the top 10\% of shots sorted by spectral intensity at the location of the weakest fringe. This filtering helped to enhance the fringe contrast. From here on we refer to the two sets of spectra as the ``unfiltered" and ``filtered" data, respectively. To isolate the fringes from the background spectrum which is mostly due to the first stage, we subtract off the average spectrum measured after the first stage. Figure~\ref{fig:spec_bgd_removal} shows this workflow for the six coarse delay points and both the unfiltered and filtered datasets. The dashed lines are the ones we ultimately use to estimate the fringe locations. We note that the particular method of background removal can influence the fringe positions, and is an inherent uncertainty of this situation in which the spectral intensity of the fringes is similar to that of the seed pulse as opposed to the case studied in simulations where the spectrum of the pulse train is much larger than that of the seed. We will refer to the residual spectra after background subtraction as difference spectra.

\begin{figure}[h!]
    \centering
    \includegraphics[width=0.75\linewidth]{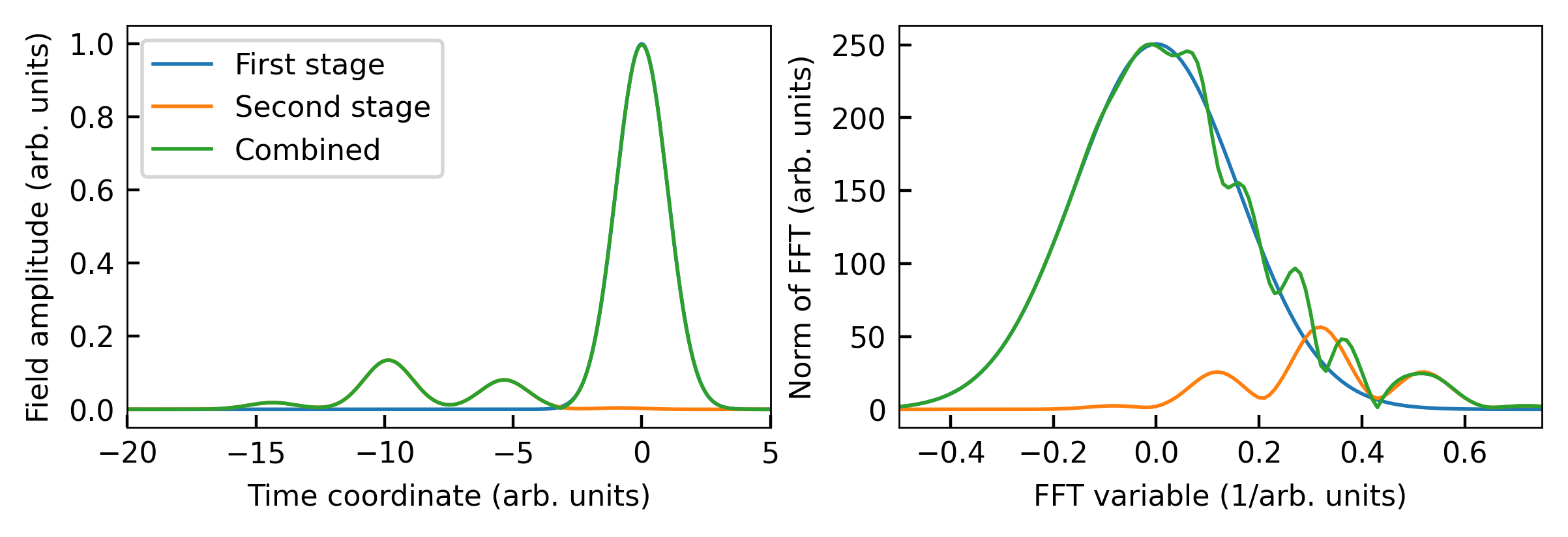}
    \caption{Toy model illustrating the structure that can appear in spectra when a pulse train is combined with a coherent gaussian pulse. }
    \label{fig:toy_model}
\end{figure}

We estimated the positions of the fringes by fitting gaussians to the difference spectra. First we estimated reasonable initial guesses for the gaussian fits by eye for the six filtered and unfiltered average spectra, then used the python package scipy to perform the fit. In the 1 fs delay dataset three fringes are clearly visible, but the distance between the two adjacent pairs appears to be different. We expect this is a result of the fact that the spectral intensities of the seed pulse and of the additional content generated in the second stage are very similar. This makes it difficult to interpret the observed fringe spacing as the fringe spacing of the isolated second stage spectrum. Figure~\ref{fig:toy_model} elucidates this concept with a toy model in which a gaussian pulse, representing the seed, is combined with a weaker pulse train. The second panel shows the spectra of the seed, the train, and the combination of the two, where it is clear that when combining the two spectra more complicated features appear depending on the precise widths and amplitudes of the envelopes and the relative phases between the seed and the pulses in the train. Those more complicated features appear in particular where the train spectrum lies on the shoulder of the gaussian spectrum, where for example the fringe spacing can appear to decrease. This explains why the left and middle fringes in the 1 fs dataset are closer together than the right and middle. For this reason, when possible we estimate the fringe spacing using the fringes farthest from the center of the seed spectrum. Note that this is not quite possible for the 1.6 fs delay data, where both fringes are on the shoulder of the gaussian and thus we expect to measure an unusually small fringe spacing.

\begin{figure}[h!]
    \centering
    \includegraphics[width=0.4\linewidth]{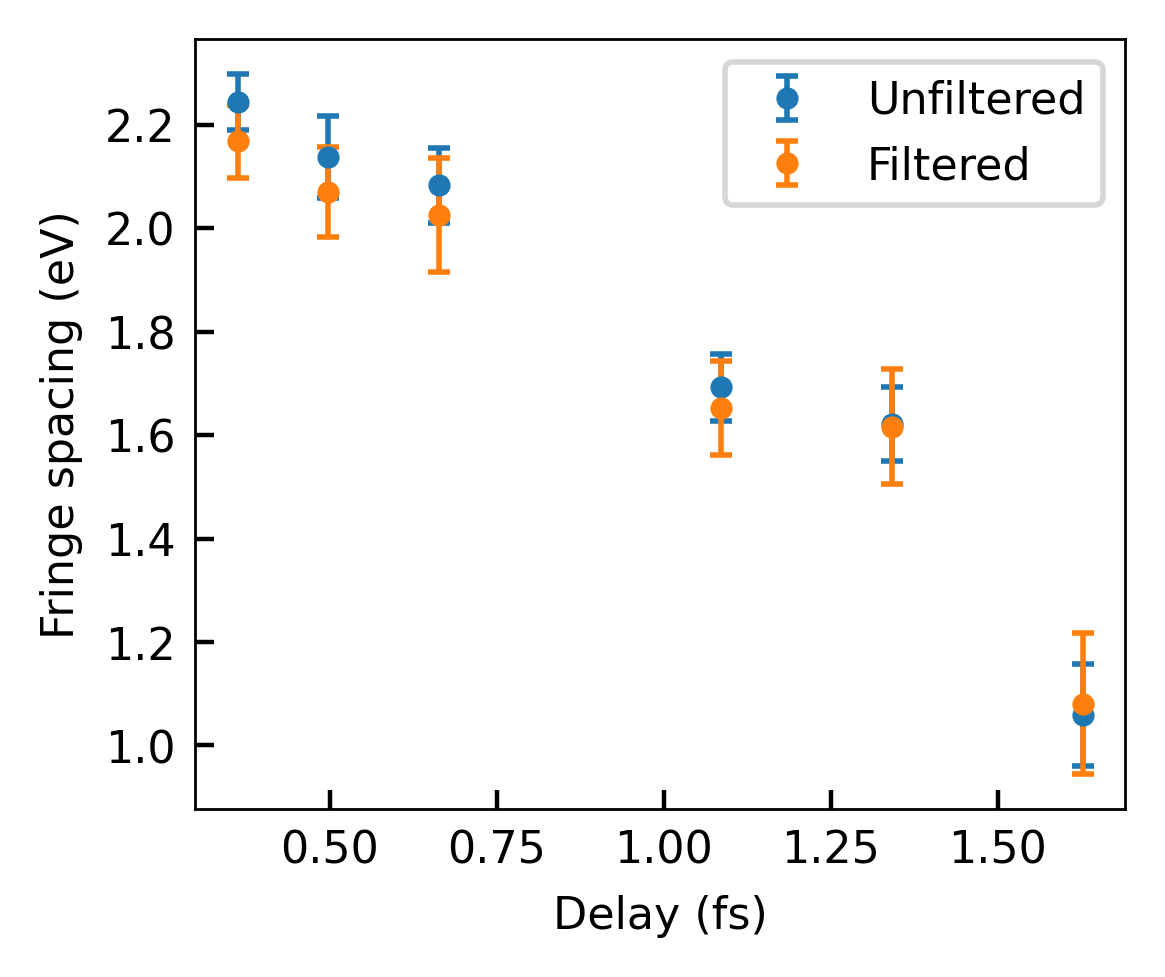}
    \caption{The estimated fringe spacing as a function of delay for the filtered and unfiltered datasets with errorbars extracted from bootstrapping with 250 iterations.}
    \label{fig:fringe_spacings}
\end{figure}

We repeated this procedure for 250 iterations of bootstrapping, where on each iteration we resampled the spectra for the average with replacement. The error bars plotted in the main paper and here are 3 times the standard deviation of the fringe spacing results for the 250 iterations. Figure~\ref{fig:fringe_spacings} shows the evaluated fringe spacing as a function of delay for the filtered and unfiltered datasets. In the main text we show only the results for the filtered dataset for clarity since the two are very close.

\begin{figure}[h!]
    \centering
    \includegraphics[width=0.4\linewidth]{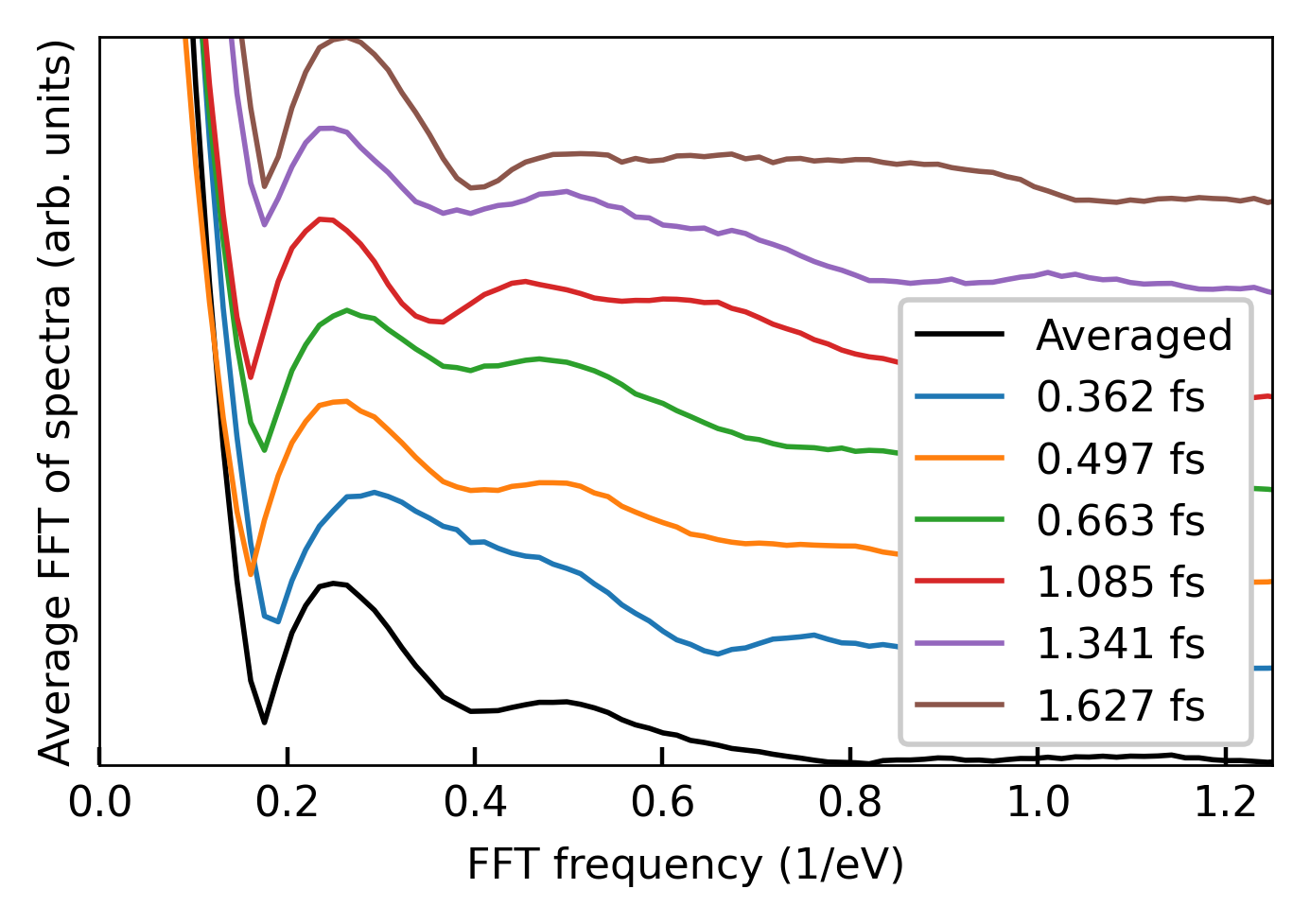}
    \includegraphics[width=0.5\linewidth]{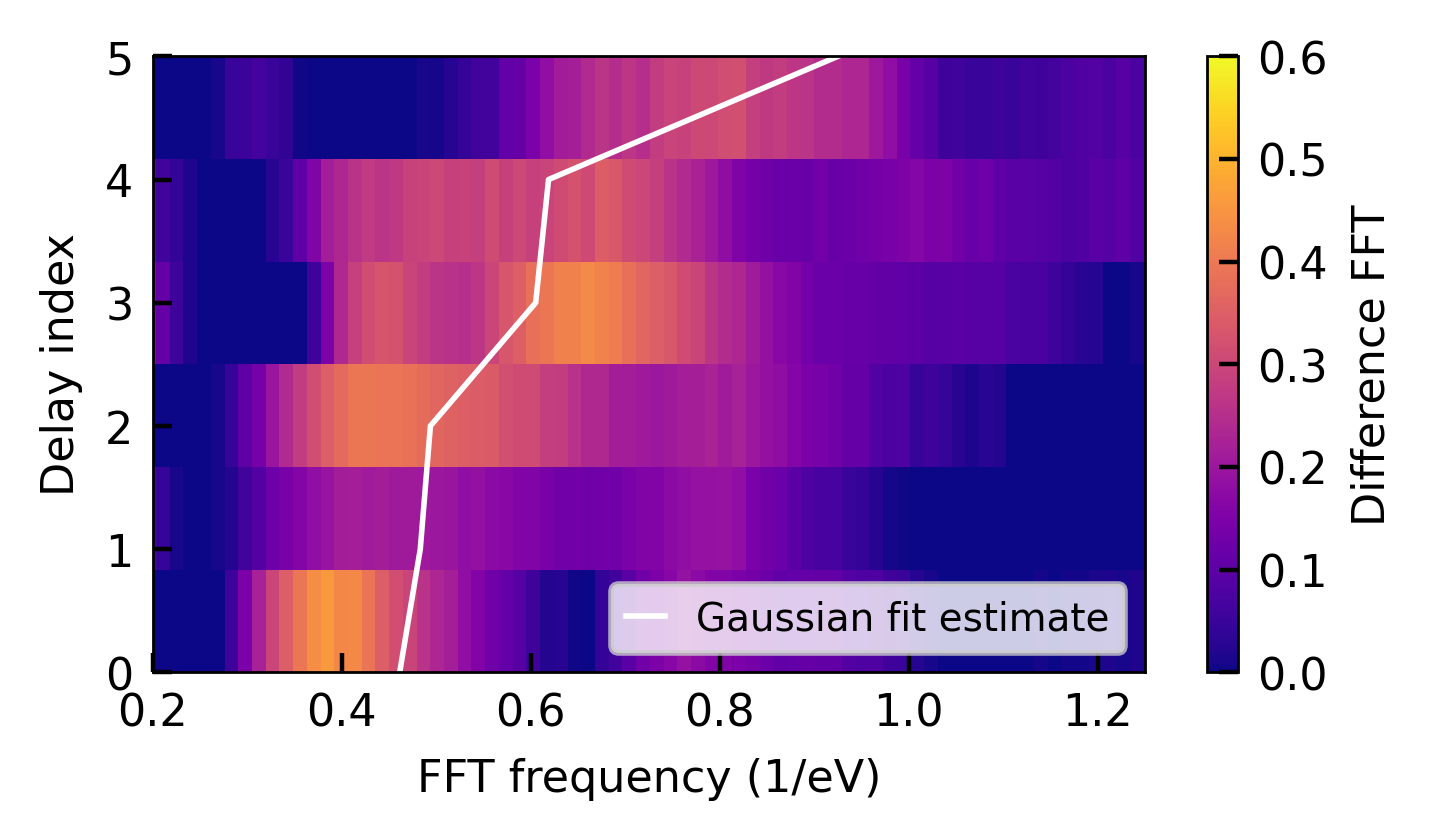}
    \caption{(Left) The FFT of the average filtered spectra for each delay in the coarse scan compared to the FFT of the average over all filtered spectra. (Right) The FFT of the average filtered spectra with the average FFT subtracted off to isolate delay specific content. The white line shows the expected position of a peak from the earlier gaussian fitting approach. }
    \label{fig:FFT_averages}
\end{figure}

An alternative way to estimate the fringe spacing is to look at the fast Fourier transform (FFT) of the spectra. Any periodic feature in the spectrum will appear as a peak in the FFT at the inverse of the spectral spacing. That analysis, like our fitting directly in the time domain, is complicated by the strong presence of the spectrum from the first stage. In addition to the FFT features deriving from the fringe spacing, there are other features due to the different widths of the seed and train envelopes, and the spacing between the seed and the pulse train. The FFTs of the average filtered spectra are shown in Figure~\ref{fig:FFT_averages} (left). Alongside the FFTs at each delay, we also show in black the FFT of the average over all of the filtered spectra, which we expect to roughly wash out features specific to any particular fringe spacing. The average spectrum shows a peak at roughly 0.25=1/(4 eV), similar to the spacing between the seed spectrum and the spectrum of the second stage, and at the harmonic of that 0.5=2/(4 eV). The other curves show a trend of having more content at higher FFT frequencies as the delay is increased, as we would expect from a shrinking fringe spacing. We expect that we can extract the particular features specific to the delay-induced fringe spacing by looking at the difference between the FFT at one delay and the average FFT. Those difference FFTs are plotted in the right panel Figure~\ref{fig:FFT_averages}, alongside the estimated peak location from our earlier gaussian fitting approach.

\begin{figure}[h!]
    \centering
    \includegraphics[width=0.5\linewidth]{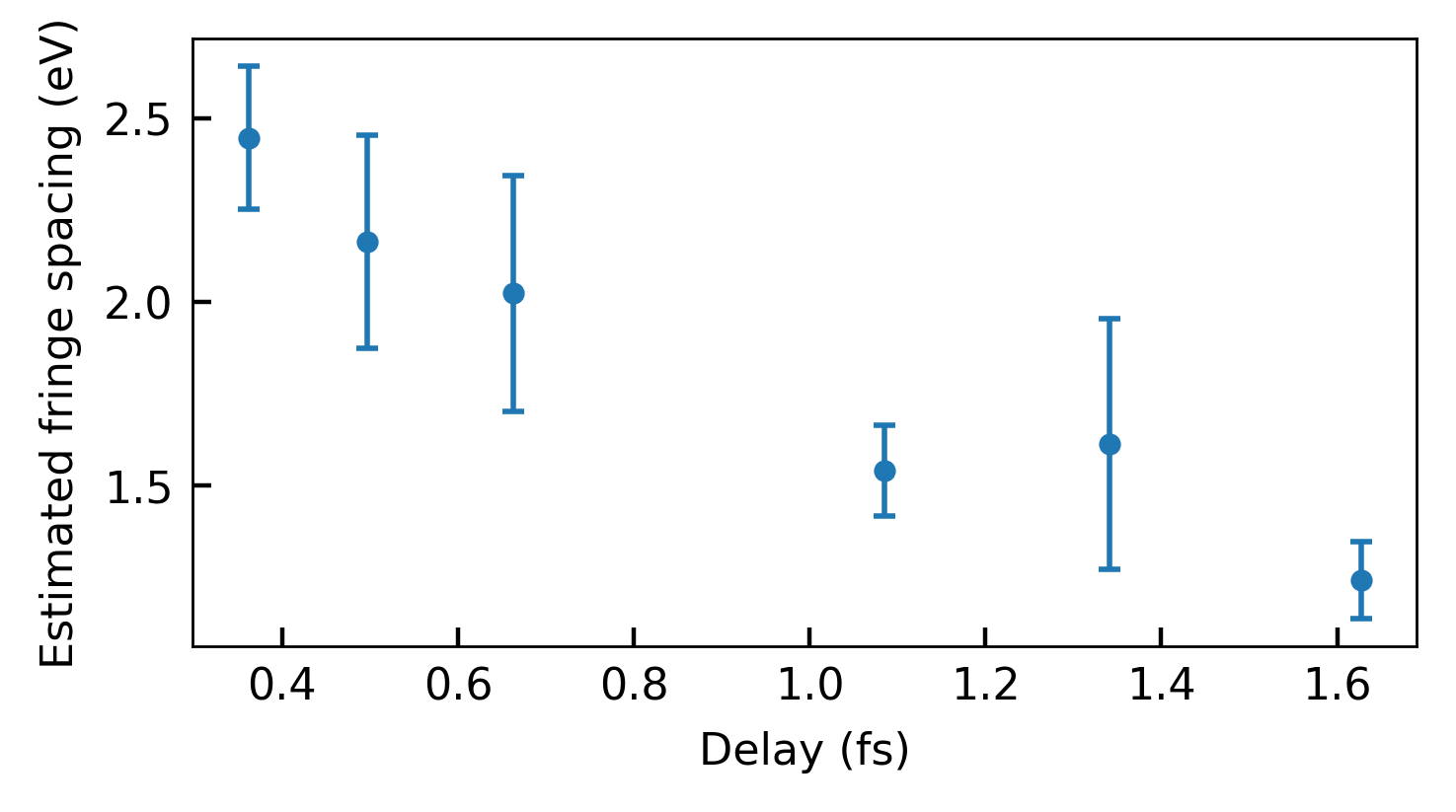}
    \caption{The estimated fringe spacing inferred from the center of mass of the difference FFTs. Error bars show three times the standard deviation of 250 bootstrapping iterations.}
    \label{fig:fft_estimate}
\end{figure}

Note that the difference FFT generally has peaks around some lower frequency and double that frequency. To estimate the position of the peak we fit a gaussian to the lower of the peaks and evaluate the center of mass of the difference FFT weighted by the fitted gaussian. We repeated this approach using bootstrapping with 250 resamplings, and show the results in Figure~\ref{fig:fft_estimate} with the error bars again indicating 3 times the standard deviation of the bootstrapped results. We observe generally larger error bars from this approach, but otherwise the same general trend and similar quantitivate values for the spacing at each delay.

\normalem 
\bibliography{apssamp}